\documentclass[12pt]{article}
\usepackage[margin=1in]{geometry}
\usepackage{setspace}
\usepackage{graphicx}
\usepackage{lscape}
\usepackage{amscd,amssymb}
\usepackage{amsmath}
\usepackage{amsthm}
\usepackage{amsfonts}
\usepackage{mathrsfs}
\usepackage{titlesec}
\usepackage{enumerate}
\usepackage{multirow}
\usepackage{subfig}
\usepackage{booktabs}
\usepackage[authoryear]{natbib}
\usepackage[usenames,dvipsnames]{color}
\usepackage{setspace} 
\usepackage[colorlinks=true,urlcolor=black,citecolor=color,linkcolor=color,bookmarks=true]{hyperref}
\usepackage{lscape,tabularx}
\usepackage{color,soul}
\usepackage{verbatim}
\usepackage{multicol, algorithm, algorithmic}
\definecolor{color}{RGB}{0,0,0}
\usepackage[flushleft]{threeparttable}
\usepackage{enumitem}
\usepackage{caption}
\usepackage{soul,framed}
\usepackage{tabu}
\usepackage{capt-of}
\usepackage{dsfont}
\usepackage{placeins}
\usepackage{datetime}
\usepackage{enumitem}
\usepackage[normalem]{ulem}

\numberwithin{equation}{section}

\usepackage{pxfonts}
\usepackage[T1]{fontenc}

\newcommand\fnote[1]{\captionsetup{font=small}\caption*{#1}}

\newcommand{\tref}[1]{\textcolor{color}{Table}~\ref{#1}}

\newcommand{\cref}[1]{Chapter~\ref{#1}}

\newcommand{\fref}[1]{\textcolor{color}{Figure}~\ref{#1}}

\newcommand{\be}[1]{\medskip{} \begin{enumerate}\itemsep=#1mm}
\newcommand{\ee}{\end{enumerate}}
\newcommand{\bq}{\medskip{}  \begin{quote}}
\newcommand{\eq}{\end{quote}}
\newcommand{\bd}[1]{\medskip{} \begin{description}\itemsep=#1mm}
\newcommand{\ed}{\end{description}}
\newcommand{\bi}[1]{\medskip{} \begin{itemize}\itemsep=#1mm}
\newcommand{\ei}{\end{itemize}}
\newcommand{\bfl}{\begin{flushleft}}
\newcommand{\efl}{\end{flushleft}}
\newcommand{\bfr}{\begin{flushright}}
\newcommand{\efr}{\end{flushright}}
\newcommand{\bc}{\begin{center}}
\newcommand{\ec}{\end{center}}
\newcommand{\bfns}{\begin{footnotesize}}
\newcommand{\efns}{\end{footnotesize}}
\newcommand{\bss}{\begin{scriptsize}}
\newcommand{\ess}{\end{scriptsize}}
\newcommand{\bns}{\begin{normalsize}}
\newcommand{\ens}{\end{normalsize}}

\usepackage{array}
\newcolumntype{R}[1]{>{\raggedleft\let\newline\\\arraybackslash\hspace{0pt}}m{#1}}

\begin{document}
\renewcommand{\topfraction}{.85}
\renewcommand{\bottomfraction}{.7}
\renewcommand{\textfraction}{.15}
\renewcommand{\floatpagefraction}{.66}
\renewcommand{\dbltopfraction}{.66}
\renewcommand{\dblfloatpagefraction}{.66}
\setcounter{topnumber}{9}
\setcounter{bottomnumber}{9}
\setcounter{totalnumber}{20}
\setcounter{dbltopnumber}{9}

\newcommand{\bdm}{\begin{displaymath}}
\newcommand{\edm}{\end{displaymath}}
\newcommand{\beq}{\begin{equation}}
\newcommand{\eeq}{\end{equation}}
\newcommand{\can}{\citeasnoun}
\newtheorem{proposition}{Proposition}
\newcommand{\authorblock}[1]{\begin{tabular}{@{}c@{}}#1\end{tabular}}

\vspace{-5cm}
\title{Equity Impacts of Dollar Store Vaccine Distribution\thanks{Authors are listed alphabetically and contributed equally. The authors thank the Tobin Center at Yale University for funding.}}
\author{Judith A. Chevalier,{\emph{Yale School of Management and NBER}}\\ 
Jason L. Schwartz, 
\emph{Yale School of Public Health}\\
Yihua Su\emph{\small{}},
\emph{Yale School of Public Health}\\ 
Kevin R. Williams\emph{\small{}},
\emph{Yale School of Management and NBER}\\
}

\date{April 2, 2021}

\maketitle

\vspace{-3ex}
\begin{abstract}
\setstretch{1}
We use geospatial data to examine the unprecedented national program currently underway in the United States to distribute and administer vaccines against COVID-19. We quantify the impact of the proposed federal partnership with the company Dollar General to serve as vaccination sites and compare vaccine access with Dollar General to the current Federal Retail Pharmacy Partnership Program. Although dollar stores have been viewed with skepticism and controversy in the policy sector, we show that, relative to the locations of the current federal program, Dollar General stores are disproportionately likely to be located in Census tracts with high social vulnerability; using these stores as vaccination sites would greatly decrease the distance to vaccines for both low-income and minority households. We consider a hypothetical alternative partnership with Dollar Tree and show that adding these stores to the vaccination program would be similarly valuable, but impact different geographic areas than the Dollar General partnership. Adding Dollar General to the current pharmacy partners greatly surpasses the goal set by the Biden administration of having 90\% of the population within 5 miles of a vaccine site. We discuss the potential benefits of leveraging these partnerships for other vaccinations, including against influenza. \\
\end{abstract}

\clearpage

\section{Introduction}

An unprecedented national program to distribute and administer vaccines against COVID-19 has been underway in the United States since December 2020. The complex storage and handling requirements of several vaccines authorized by the U.S. Food and Drug Administration (FDA), as well as the limited vaccine supply available in the initial months of the vaccination campaign, led public health officials to create dedicated vaccination sites largely outside of the clinical settings where individuals typically receive medical care. These locations---a mix of federal-, state-, and locally-run sites of widely varying size and located in diverse settings---are intended to facilitate the rapid administration of available vaccine doses while being attentive to the goal of promoting equity in vaccine delivery \citep{bibbins2021taking}. To address this latter objective, health officials have pursued strategies to enhance vaccine access and availability in communities that have been disproportionately affected by COVID-19 and its most severe outcomes, including low-income and minority communities that have long encountered barriers to the receipt of necessary health care services, including vaccination.

In March 2021, the director of the U.S. Centers for Disease Control and Prevention (CDC) and the company Dollar General confirmed reports that they were exploring a partnership through which COVID-19 vaccines would be administered in Dollar General locations \citep{dollargeneralwebsite,dollargeneralUSA}. This potential partnership is particularly noteworthy because the growth of the two national dollar store chains, Dollar General and Dollar Tree, has generated considerable skepticism and controversy in the policy sector. For example, citing arguments that the growth of dollar stores in lower-income areas contribute to food insecurity, numerous localities have passed ordinances restricting the growth of these chains \citep{cnndollar,axiosdollar}.  Despite this controversial context, if an agreement were to be reached, this partnership would augment the federal government's existing partnership with retail pharmacies. Although such a partnership would likely not ultimately involve all 16,805 Dollar General locations, a partnership with the chain has the potential to greatly increase the scope of the current Federal Retail Pharmacy Partnership Program. 

Under the current scheme for distributing COVID-19 vaccines in the United States, the federal government controls all vaccine allocation. On a weekly basis, nearly all available doses are allocated to state governments or directly to participants in the Federal Retail Pharmacy Partnership Program.\footnote{A small number are reserved for direct federal vaccination efforts, such as those coordinated by the Department of Defense or Department of Veterans Affairs, as we describe in more detail elsewhere in this paper.}  The CDC describes the Federal Retail Pharmacy Program for COVID-19 Vaccination as "a collaboration between the federal government, states and territories, and 21 national pharmacy partners and independent pharmacy networks to increase access to COVID-19 vaccination across the United States" \citep{CDCpharm}. The program is intended to work in parallel with vaccination sites established and supported at the state level (and using state-allocated vaccine doses), thereby advancing the federal government’s objective of ensuring broad vaccine access. The program is being implemented and expanded incrementally based on available vaccine supply and as needs are identified in particular communities, with select retail pharmacy locations providing COVID-19 vaccine to individuals eligible for vaccine based on state prioritization guidelines.

This type of partnership between the federal government and retail pharmacies is novel, but it reflects the expanding role of pharmacies in providing clinical services to patients in recent years, including the administration of recommended vaccines \citep{cassel2018can}. For COVID-19 vaccination, the pharmacy partner locations are seen as particularly valuable toward reducing the distance required for Americans to reach their nearest vaccination site. To this end, the Biden administration has announced its goal of having a vaccination site located within 5 miles of 90\% of American adults by April 19, 2021 \citep{WHfactsheet}.

In this paper, we use geospatial data to examine the proximity of U.S. households to their closest federal pharmacy program partner and quantify differences in access based on geographic data. We show that, relative to the locations of current federal pharmacy partner stores, Dollar General stores are disproportionately likely to be located in Census tracts with high social vulnerability, as measured using the Center for Disease Control's Social Vulnerability Index 2018 Database US (SVI), accessed March 17, 2021 \citep{CDCSVIB}. As low household income is an important component of social vulnerability, we further show that adding Dollar General as a partner would dramatically improve the fraction of low-income households that live less than a mile from their closest vaccination program partner, particularly in the South and Southeast, compared to the current retail pharmacy program. Thus, the Dollar General partnership  would have the effect of greatly expanding vaccine access for these groups. We also show that adding Dollar General as a partner disproportionately improves access for Black individuals and for Hispanic and Latino people. For individuals at all income levels, full utilization of existing pharmacy partner sites  would approach the proximity target (90\% within 5 miles) set by the Biden administration; adding Dollar General to the vaccination program would greatly surpass that goal. 

Because of the strong results identified with the proposed partnership with Dollar General, we also examine a hypothetical partnership with Dollar Tree---the other major dollar store chain in the United States. We show that Dollar Tree stores are even more concentrated in locations with high social vulnerability than are Dollar General stores. For some states, adding Dollar Tree stores to the Federal Retail Pharmacy Partnership Program would provide a markedly better reduction in the proximity of low-income households to their closest federal vaccination partner site than would adding Dollar General to the program. However, for some states, the reverse is true.  

Finally, we recognize that the Federal Retail Pharmacy Partnership Program is only one channel of vaccine access. For a subset of states (20), we gather all active vaccination locations in that state, as listed on that state's COVID-19 response website. We consider the superset of sites from the federal pharmacy program and sites set up by the states. Several states include mass vaccination locations, federal pharmacy partners, and federally qualified health centers in their directories. We show that adding state-listed vaccination sites to our baseline analysis of the federal pharmacy program increases the share of low-income households within 1 mile or 5 miles of a vaccine site. However, we show that adding Dollar General stores nonetheless provides a substantial proximity improvement for low-income households in many states.

Our paper proceeds as follows. Section 2 discusses our data and methodology. In Section 3, we profile the major dollar store chains and document their recent growth. In Section 4, we describe the retail pharmacy program and quantify the characteristics of Census tracts in which retail pharmacy partners are located. In Section 5, we quantify the impacts of adding dollar stores to the federal vaccine program on vaccine access. In Section 6, we undertake a partial analysis of state vaccine locations. Section 7 concludes.

\section{Data and Methodology}

Our purpose is to assess the characteristics of partners in the Federal Retail Pharmacy Partnership Program and Dollar Stores and quantify their proximity to populations identified as high social vulnerability. We are especially interested in low-income populations. 

We bring together data from several sources.  First, we obtain a list by state of federal retail pharmacy partner chains from the Centers for Disease Control \citep{CDCpharm}. As described by the CDC, the program is comprised of "21 national pharmacy partners and independent pharmacy networks."  In providing a list of pharmacy partners for each state, the separation of the pharmacy efforts and state efforts is emphasized. The CDC notes, "This list is specific to the Federal Retail Pharmacy Program; some states have engaged additional pharmacies directly to assist with vaccination efforts". 

For each pharmacy partner, in each state, we match these chains by name to the Historical Business Information Files from ReferenceUSA \citep{ReferenceUSA}, accessed March 14, 2021. We were unable to match the pharmacy network CPESN, a partner in eleven states, and MHCN, a partner in four states. From ReferenceUSA, we obtain the universe of locations of these federal partners.  

It is important to note that not all pharmacy locations for a given chain (and state) are necessarily providing vaccines. Moreover, some locations may be providing vaccinations intermittently. This is because the supply of vaccine has been constrained overall, but also because some retailers---due to freezer constraints, for example---can only offer a subset of the vaccines. Thus, some chains only offer vaccinations when specific vaccines are available. We nonetheless include all of the outlets of these chains because the number of locations offering vaccinations is expected to increase as vaccine supply grows. In March 2021, representatives of these chains explicitly noted that more of their outlets will provide vaccines as supply increases, a plan confirmed by the Biden administration \citep{barker,robbins,WHfactsheet}.

We further supplement our analysis using vaccine locations found on state COVID response websites. State vaccine sites are dynamic; we use active vaccine sites as of the week of March 22rd, 2021. Therefore, our analysis does not necessarily reflect the complete buildout of state capacity, although we do include in our analysis announced locations that did not have appointments available at the time of data collection. Some states also list pharmacies found through the federal partnership program; our analysis is unaffected by potential duplicates as we select the closest vaccination site to each Census tract. Additionally, these files may allow us to incorporate Federally Qualified Health Centers and locations that we cannot match using the ReferenceUSA data.

We obtain vaccine locations for 21 states that either post a file containing all locations or present users with an interactive map of locations.\footnote{We collect data for AL, AR, CT, GA, IL, KS, MD, ME, MO, MS, ND, NJ, NV, OH, OK, PA, SC, WA, WI, and WV. First, we download either the PDF, HTML, or JSON found on from each state's COVID-19 response website. We then process the files and extract the addresses. We then geocode each site using https://geocode.localfocus.nl/. Our analysis includes "success" and "doubt" matches, and excludes "failed" matches.} \fref{state_links} contains a list of the states and the corresponding websites we used for data collection.

In order to examine the implications of a partnership with Dollar General and a hypothetical partnership with the other large dollar store chain, Dollar Tree, we download all of their locations from ReferenceUSA. For all retail dollar store and pharmaceutical outlets, we remove a small number of outlets that are listed in ReferenceUSA from analysis if they appear to be a headquarters, a distribution center, or other non-retail location of the company.   

Our goal is to examine the social vulnerability metrics for the Census tracts containing these outlets and measure distances from the these retail outlets to households.  We use the latitudes and longitudes of the retail outlets provided by ReferenceUSA and map these to Census data on households. Throughout this analysis, we will use Census data at the Census tract level. There are approximately 74 thousand Census tracts in the US, created by the Census Bureau with a target population of 4,000 per tract. Some tracts were unavailable for processing, leaving us with 73,088 Census tracts. As is common in the literature, we will assume that all households live at the geographic centroid of the Census tract and calculate the great circle distance from the Census tract centroids to each of the retail outlets in our database.  We use the minimum-calculated distance for each Census tract to a vaccine location for our analysis. Because vaccine eligibility is set by individual states and allocations are at the state level, we calculate distances to the closest same-state retail location; we assume that people located near a state border will not obtain the vaccine at an out-of-state location. \footnote{We make an exception for North Dakota, for which we have not found retail outlets of any federal pharmacy partner that was listed as a partner for North Dakota on the CDC website.} 

We also characterize the Census tracts within which the retail outlets are located.\footnote{We use the 2019  American Community Survey 5-Year Data (2009-2019), Census Bureau Geocoder, and TIGER/Line Shapefiles downloaded from \\ { https://www.baruch.cuny.edu/confluence/display/geoportal/US+Census+Population+Centroids}} To do this, we match each Census tract not only to the Census data provided directly by the Census, but to the tract-level data on the Social Vulnerability Index 2018 Database US (SVI) provided by the CDC. SVI is a composite of Census data used to identify communities that may require the greatest support during, or following, a disaster; it has been widely adopted by federal and state health officials during COVID-19 as a tool to inform the design of vaccination efforts and to assess their performance with respect to equity \citep{CDCSVIB,hughes2021county}. We are able to match 72,173 SVI tracts to the Census data.



In our examination of the current composition of the pharmacy partner program, we find 26,246 retail pharmacy outlets that belong to chains that are designated as pharmacy partners in the state in which the outlet is located. Only a subset of these outlets are currently administering vaccinations, but some of these chains will significantly expand the number of outlets within the chain administering vaccinations as vaccine supply increases. In matching to Census data, we summarize in Table \ref{tab:summarypharm} the proximity of various segments of the population to the closest within-state outlet of the Federal Retail Pharmacy Partner Program. We examine income groups, racial groups, and Hispanic/Latino ethnicity.\footnote{To characterize race, we divide the total population into one of four racial groups: white (for which we use the Census categorization "White alone"), Black (for which we use the Census categorization "Black or African American", AAPI (for which we add the Census categorizations of Asian American, Native Hawaiian and Pacific Islander), and other races for which we include all other races. We also characterize the population as "Hispanic" or "non-Hispanic", including the Census categorizations of Hispanic or Latino and including Hispanic or Latino people of all races.}  


\begin{table}
 \setlength{\tabcolsep}{10pt}
 \centering
\caption{Characteristics of U.S. Pharmacy Partner Program \label{tab:summarypharm}}
\begin{threeparttable}
\begin{tabular}{  l  l c  c  c  }
\toprule
Population Type &  Area  &  Share  &  Share   & Share  \\
  &    &  $<$1 mile &  $<$2 miles &  $<$ 5 miles   \\ 
  \midrule
All adults & U.S. & 48.3 & 	72.7 & 	86.3  \\
All adults & CONUS  & 48.7  & 	73.0 & 	86.4   \\
Households $<$ \$35K & U.S. & 48.4 & 71.1 & 83.3  \\
Households $>$ \$100K & U.S. &  49.3 & 	73.9 & 87.8  \\
Black Population & U.S.  & 53.6 & 81.8 & 	93.2  \\
White Population & U.S. & 44.3 & 68.6 & 83.8  \\
AAPI Population & U.S. & 65.6 & 90.3 & 97.5  \\
Other Population & U.S. & 58.5 & 80.7 & 90.4  \\
Hispanic Population & U.S. & 44.9 & 70.0 & 84.6 \\
non-Hispanic Population & U.S. & 61.2 & 84.2 & 93.7 \\

\bottomrule
\end{tabular}
\begin{tablenotes}
\item Notes: Summary statistics for federal U.S. pharmacy partner program. Table entries denote the share of the population type within the denoted proximity to an in-state pharmacy partner. "CONUS" denotes continental US.  
\end{tablenotes}
\end{threeparttable}
\end{table}

Table \ref{tab:summarypharm} shows that these pharmacy partners (if all outlets were providing vaccine) are nearly sufficient to meet the Biden administration's promise that 90 percent of Americans will be within 5 miles of a vaccine site by April 19, 2021. As this chart includes only the federal pharmacy program vaccine sites (and we were unable to match a small number of pharmacies), providing vaccines at all of these sites would almost surely be sufficient to achieve this pledge. Nonetheless, the administration pledge is not the only way to measure access. In particular, the table shows that the fraction of households within 1, 2, or 5 miles of a federal pharmacy partner is smaller for low-income households than for high-income households. This is concerning because low-income households are least likely to have access to transportation infrastructure, internet access to book a distant appointment, etc. That the majority of these households are more than a mile from the closest pharmacy partner could indeed create a barrier to vaccination. The table also shows a large disparity in proximity between Hispanic and non-Hispanic people. This disparity in proximity is particularly concerning as COVID-19 vaccination rates for Hispanic people have been shown to be low. \citep{KFFdisparities} An important caveat to our analysis is that we consider the universe of outlets in Federal Retail Pharmacy Program chains. As discussed earlier, not all chain pharmacy locations have been used to distribute vaccines. Adding or removing locations can increase or reduce disparities in vaccine access.

We obtain addresses for 10,439 vaccination sites on state websites, corresponding to 8,455 unique geographic coordinates. The mean and median number of locations per state is 403 and 275, respectively. There are a few outliers. The West Virginia file contains only 23 locations because it lists only state-coordinated vaccine clinics. Missouri and Ohio list over 1,300 locations.\footnote{The website https://covidvaccine.mo.gov/map/ shows 1,639 vaccination websites, accessed 3/31/2021. The website https://coronavirus.ohio.gov/wps/portal/gov/covid-19/dashboards/covid-19-vaccine/covid-19-vaccine-provider-dashboard lists 1,332, accessed 3/31/2021. }

\section{Dollar Store Chains in the United States}
\begin{table}
 \setlength{\tabcolsep}{10pt}
 \centering
\caption{Summary statistics for U.S. Dollar Stores, selected years \label{tab:summarytable}}
\begin{threeparttable}
\begin{tabular}{  l  l l l  }
\toprule
	 &  & No.   \\ 
	Year \hspace{.75in} & Chain \hspace{1.75in}  & Stores    \\ \toprule
	    & &  \\
	2005 & Dollar General & 7,086   \\ 
	& Family Dollar & 5,406     \\ 
	& Dollar Tree & 2,654      \\ 
	&  &   \\
	2012 & Dollar General & 9,617    \\ 
 & 	Family Dollar & 7,257   \\ 
 & 	Dollar Tree & 4,160     \\ 
 &  &    \\
	2019 & Dollar General & 15,317    \\ 
 & 	Dollar Tree/Family Dollar & 14,495   \\ 
 &  &       \\ 
	2021 &  Dollar General & 16,805   \\ 
 & 	Dollar Tree/Family Dollar & 15,629    \\ 
 \bottomrule
\end{tabular}
\begin{tablenotes}
\item Notes: Number of outlets of Dollar Tree and Family Dollar by year. Dollar Tree and Family Dollar merged in 2015. Some stores owned by Dollar Tree are still branded Family Dollar in 2021. Data from RefUSA.   
\end{tablenotes}
\end{threeparttable}
\end{table}

The major dollar store chains have expanded dramatically in the United States over the last fifteen years. As \tref{tab:summarytable} demonstrates, the number of outlets of the major dollar store chains have more than doubled since 2005. As demonstrated by the 2019 and 2021 store counts, substantial growth in stores, particularly Dollar General, continued even during the COVID-19 pandemic.

\begin{figure}[!ht]
\caption{Dollar General and Dollar Tree Locations; Stores per 100K Population}
\begin{center}
\includegraphics[trim={0 .15cm 0 .15cm},width=172mm]{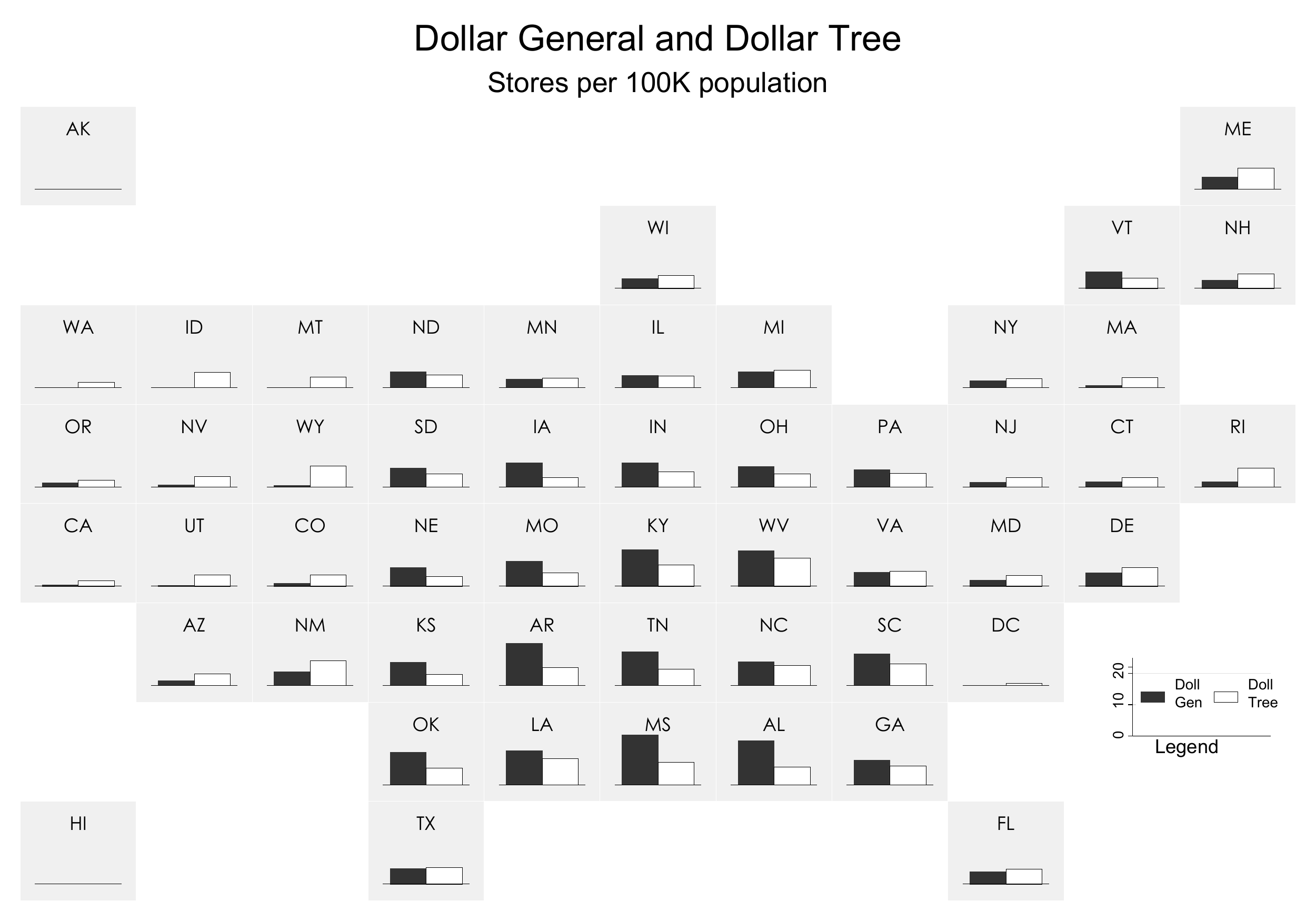}
\end{center}
\label{fig:DGDTmap} 
     \fnote{\small  Notes: Depiction of Dollar General and Dollar Tree stores per 100 thousand population by state. Dollar General is represented by the black bar on the left and Dollar Tree by the white bar on the right. Raw data are shown in \fref{tab:dollarstores} in the Appendix.}
\end{figure}

While dollar stores blanket the continental US, the coverage of the two major chains is particularly dense in the Southeast area of the country. Even a cursory inspection of their prevalence by state demonstrates that they are particularly numerous in several of the poorest states in the United States. As Figure \ref{fig:DGDTmap} shows, Dollar General is most dense per capita in Alabama, Arkansas, and Mississippi, with nearly 20 stores per 100 thousand population. There is also substantial density also in Kentucky, Louisiana, Oklahoma, South Carolina, Tennessee, and West Virginia. The other major dollar store chain, Dollar Tree, while similar in terms of overall store counts, is substantially more dispersed geographically. As shown in Figure \ref{fig:DGDTmap}, it is most dense in Louisiana, New Mexico, and West Virginia, with nine to ten stores per 100 thousand population. The state-by-state data underpinning the figure is in \fref{tab:dollarstores} in the Appendix.



As noted above, the rapid expansion of dollar stores in the United States has been controversial. One recent report from the Institute for Local Self-Reliance refers to dollar store chains as "an invasive species in America's left behind places" (\cite{donahue2018dollar}). In particular, recent research examines linkages between the variety of food sold at dollar stores and the prevalence of dollar stores and obesity, although some research has questioned whether there is a causal link from dollar stores to obesity directly \citep{allcott2019food}. While a causal link has not been established, some authors have argued that the influx of dollar stores cause the exit of traditional supermarkets and grocery stores, creating food deserts, and numerous localities have instituted regulations to curb dollar store growth. 

Whether the growth of dollar stores has negative impacts is beyond the scope of this research. However, our hypothesis is that the criticisms of dollar stores derive precisely from their potential advantage in a vaccine distribution program. If dollar stores are more prevalent in locations proximate to low-income households than are other types of retailers, then they are uniquely suited to improving vaccine access.  

\section{The Retail Pharmacy Program and Vaccine Access}

Since the inception of the COVID-19 vaccination program in December 2020, all vaccine doses available in the United States have been procured and distributed through the federal government \citep{HHSVaxOverview}. Each week, manufacturers of FDA-authorized vaccines---three, as of April 2021---report the number of doses available for distribution. Federal health officials reserve a portion of those doses for federal vaccination efforts, such as the Department of Defense, the Department of Veterans Affairs, and the Bureau of Prisons, as well as a small emergency reserve. Doses are then allocated through two principal channels. \footnote{In the initial months of the vaccination program, doses were also directly sent by the federal government to a program that administered vaccines to residents and staff of long-term care facilities. That effort, the Pharmacy Partnership for Long-Term Care Program, involved three pharmacy partners---CVS, Walgreens, and Managed Health Care Associates---all of whom are among the participants in the current retail pharmacy program \citep{CDCLTC}.} Most are directed to states, on a pro rata basis according to their share of the U.S. adult population. Based on those weekly allocations, state health officials place orders for quantities of vaccine doses (and vaccine types) to be delivered to specific vaccination sites organized by local health systems, local health departments, or other organizations working in partnership with that state. 

Additional doses are directed to the Federal Retail Pharmacy Program. Currently, doses are allocated to pharmacy partners in each CDC jurisdiction---the states, territories, and large cities that have relationships with CDC---on the basis of population, thereby bringing additional vaccine availability to states indirectly. The retail pharmacy partners serving a given jurisdiction determine which of their retail locations will receive vaccine doses; the CDC's description of the program states that pharmacy partners make those determination based on factors including "including equitable access, demand, supply, and market saturation" \citep{CDCLTC}.

The Federal Retail Pharmacy Program was created to ensure wide access to vaccines, complementing and expanding upon state-supported vaccination sites. Throughout the design and implementation of the vaccination effort to date, health officials have emphasized that the success of the program would be judged not simply on the number of doses administered, the percent of the population vaccinated, or the reduction in cases or severe outcomes following widespread vaccination, but also on the degree to which vaccine administration occurs equitably within and among communities. 

To inform this work on vaccine access and to evaluate its success, federal and state health officials have utilized the Social Vulnerability Index (SVI). Created by the Centers for Disease Control's Agency for Toxic Substances and Disease Registry (CDC-ATSDR) in 2011 and updated every two years, the measure is intended to capture "the degree to which a community exhibits certain social conditions, including high poverty, low percentage of vehicle access, or crowded households may affect that community’s ability to prevent human suffering and financial loss in the event of disaster. These factors describe a community’s social vulnerability."\citep{CDCSVIdocumentation}. 

The SVI is calculated by the CDC using Census data at the Census tract level.  For each Census tract, the overall SVI index (called RPL-THEMES) "ranks the tract on 15 social factors, including unemployment, minority status, and disability...."\citep{CDCSVIdocumentation}. The SVI index for a Census tract is a measure from zero to one which reports the fraction of Census tracts that have an overall social vulnerability less than the Census tract at issue.  

The stated intent of the SVI is to assist health officials in identifying those communities that may---as a result of these characteristics---require additional support during a public health emergency or other hazardous event. It has been used by states and private entities to calibrate disaster response prior to the COVID-19 pandemic \citep{flanagan2018measuring} and SVI measures have been shown to be correlated with worse COVID-19 outcomes \citep{karaye2020impact}. During the ongoing vaccination program, CDC and states have released data evaluation patterns between vaccination coverage and social vulnerability as measured via SVI \citep{hughes2021county}.

We examine the distribution of SVI measures of Census tracts in which federal pharmacy partners are located.  We compare these to the distribution of SVI measures for Dollar General Stores.  For further comparison, we also compare these to the distribution of SVI measures for the other major dollar store chain, Dollar Tree/Family Dollar.  

For the pharmacy partners and Dollar General stores in tracts with SVI designations, we look at the distribution by decile of SVI. The numerical value of SVI for Census tract $i$ is the fraction of all Census tracts which are less vulnerable than tract $i$. Thus, higher value of SVI is more vulnerable, and one tenth of all Census tracts are assigned to each SVI decile bin. In the figure, we show the fraction of all federal pharmacy partner outlets located in each decile bin of SVI, the fraction of all Dollar General stores located in each decile bin of SVI, and the fraction of all Dollar Tree stores located in each decile bin of SVI.  

The roughly 26 thousand pharmacy partners that we have mapped are nearly evenly distributed across the SVI deciles. However, the highest SVI decile, composed of the Census tracts estimated to have the highest social vulnerability, contain the smallest fraction of the federal pharmacy partner locations. The figure for Dollar General illustrates why it could be a valuable federal partner in reaching vulnerable communities. Dollar General stores are noticeably underrepresented in the lowest SVI Census tracts.

\begin{figure}[!ht]
\caption{{Distribution of stores by Census tract SVI}}
\begin{center}
\includegraphics[trim={0 .15cm 0 .15cm},width=142mm]{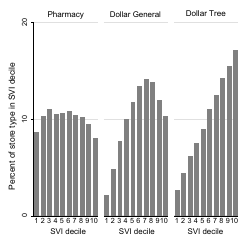}
\end{center}
\label{fig:SVIdecile} 
     \fnote{\small  Notes: Distribution of Pharmacy partners, Dollar General, and Dollar Tree Stores by social vulnerability index (SVI) deciles, continental United States. The figure shows the share of current pharmacy partners, Dollar General stores, and Dollar Tree stores by SVI decile. For example, if an outlet of a chain is located in a Census tract for which 95 percent of all Census tracts are less vulnerable using the full SVI index, the outlet will be counted to be in the top SVI decile in the graph.}
\end{figure}

\section{Evidence on the Addition of Dollar Stores to Vaccine Access}

The Social Vulnerability Index has several distinct components. An important component is household income. It is well-established that seasonal flu vaccination rates increase with income \citep{linn2010disparities}; lower-income households face numerous barriers to vaccination such as access to scheduling technologies and transportation \citep{schmid2017barriers,press2021inequities}. Vaccine take-up will likely improve with greater proximity to a vaccination site, especially a vaccination site within walking distance. Thus, we examine here the proximity to retail pharmacy outlets of low-income households. Following the Census, we consider low-income households to be households with less than \$35 thousand in annual income. We consider the proximity of the current federal pharmacy partners to these households, as well as the proximity that would occur if all Dollar General locations were added as vaccine locations.

\begin{figure}[!ht]
\caption{{Household proximity to closest pharmacy partner}}
\begin{center}
\includegraphics[trim={0 .15cm 0 .15cm},width=142mm]{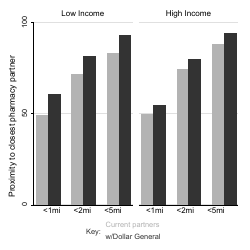}
\end{center}
\label{fig:distlowincome} 
     \fnote{\small  Notes: The graph shows the share of households within different proximity bands to the current pharmacy partners (the grey bars) and the current pharmacy partners plus Dollar General (the black bars). The left panel shows low-income households in the continental United States and the right panel shows high-income households.}
\end{figure}

As Figure \ref{fig:distlowincome} demonstrates, in the continental United States, 48.9 percent of low-income households reside within one mile of an outlet of one of the current pharmacy partner chains. This would represent an overestimate of access from this program because, as discussed above, not all outlets of these chains are or will supply vaccines. This climbs to 60.5 percent if the vaccine were offered at all Dollar General outlets. For comparison, the right panel of Figure \ref{fig:distlowincome} provides comparable statistics for high-income households with incomes greater than \$100 thousand. A slightly higher fraction of high-income households are close to a federal pharmacy site than are low-income households, 49.5 percent. However, as might be expected by the high average SVI index of dollar store locations, adding Dollar General as a partner does not increase the fraction of households located within one mile of an outlet as dramatically as it does for low-income households. While adding Dollar General to the vaccine distribution program increases the share of low-income households within a mile of a partner from 48.9 percent to 60.5 percent, for high-income households the share within a mile of a partner is increased from 49.5 percent to 54.9 percent with the addition of Dollar General locations. Similarly, when considering less than 2 mile proximity and less than 5 mile proximity, a smaller share of low-income households have proximity to pharmacy partners than do high-income households for each cutoff. Adding Dollar General reverses this, and a larger fraction of low-income households are within 2 or 5 miles of a partner when Dollar General is added to the program.  

Currently, we estimate that 86.3 percent of U.S. adults live within 5 miles of an outlet of one of the federal pharmacy partner chains. Adding Dollar General would raise that to 94.3 percent of U.S. adults living within 5 miles of at least one outlet of a pharmacy partner chain or Dollar General. This accounting does not include state and local vaccination sites, suggesting that there are multiple routes to achieving the Biden administration's goal of 90 percent of adults residing within 5 miles of a vaccination site. 

\begin{figure}[!ht]
\caption{Percentage of Low-Income Households with Federal Pharmacy Partner at <1mi}
\begin{center}
\includegraphics[trim={0 .15cm 0 .15cm},width=172mm]{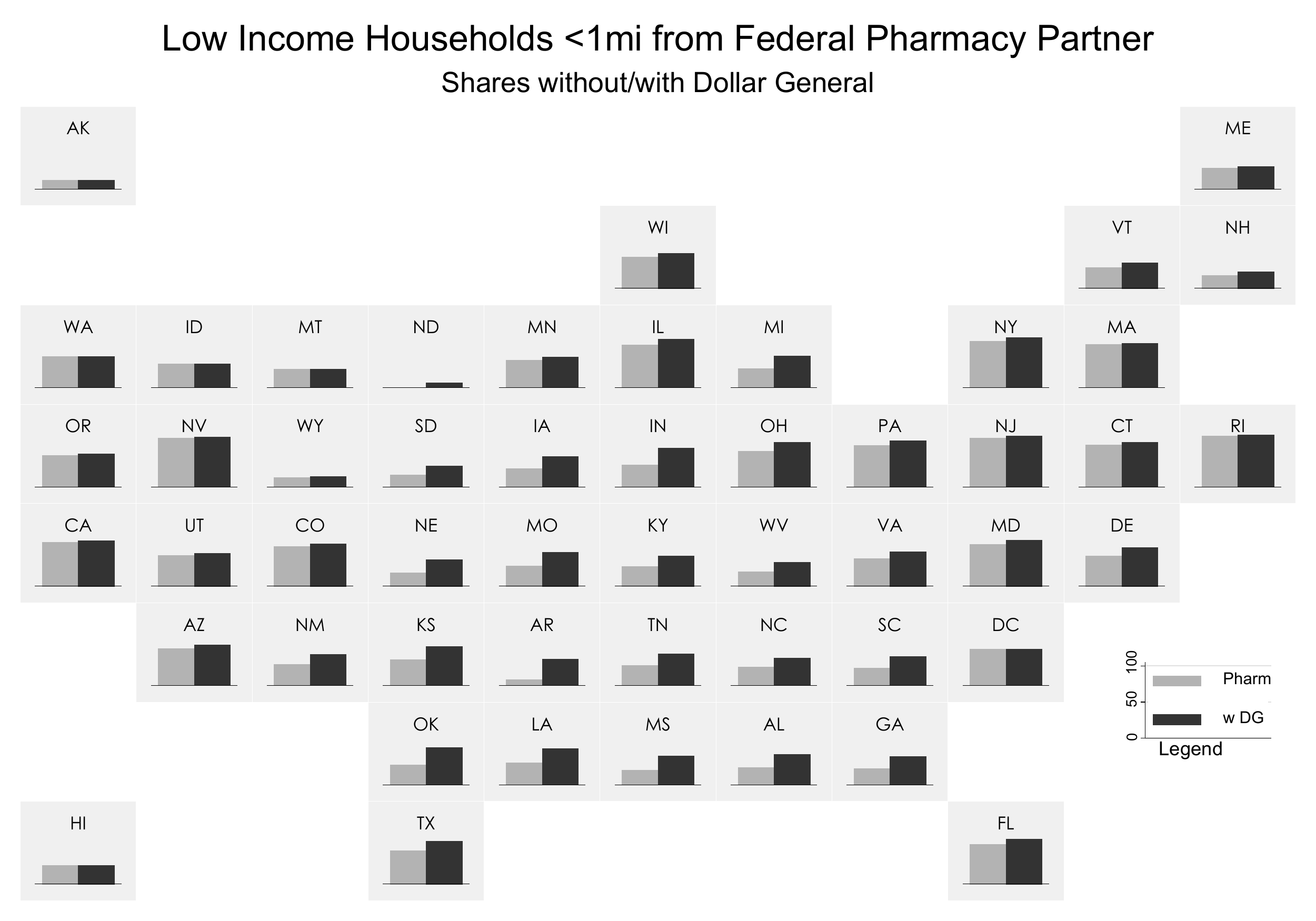}
\end{center}
\label{fig:distimprovementstate} 
     \fnote{\small  Notes: State-by-state data on the share of households earning less than \$35K per year that are located less than a mile from a federal pharmacy partner. The grey bar represents the current pharmacy partners and the black bars add Dollar General as a partner. }
\end{figure}

While the overall continental U.S. pattern illustrates the substantial proximity improvements of adding Dollar General as a pharmacy partner, the overall data masks considerable cross-state heterogeneity.  Figure \ref{fig:distimprovementstate} focuses on the less than one mile distance. The figure shows, for each state, the share of low-income households located less than one mile from a federal pharmacy partner. On the left of each bar pair for each state, the current pharmacy partner is shown and on the right the impact of adding Dollar General is illustrated. In some states, such as Alaska and Hawaii, there are no Dollar General stores and thus the figure shows no improvement (the two bars are of equal height). However, in some states, particularly in the South and Midwest, the hypothetical addition of Dollar General to the pharmacy program dramatically increases the share of low-income households less than one mile from a pharmacy partner. The raw data are included in \fref{tab:pharm_inc_tbl} in the Appendix.

Racial and ethic disparities in COVID-19 vaccine distribution have been identified. White and Asian individuals have been vaccinated at a rate greater than or equal to their share of the population, share of cases, and share of deaths. In contrast, Black and Hispanic individuals had, as of the end of March, been undervaccinated relative to their prevalence in the population and to their share of COVID-19 cases and share of deaths \citep{KFFdisparities}. Improving proximity of vaccine providers to these groups may help ameliorate such disparities. To explore this, we measure the impact of adding Dollar General to the Federal Retail Pharmacy Partnership Program on proximity of vaccination cites to different racial and ethnic subgroups of the population. These are shown in Table \ref{tab:summaryrace}. 

\begin{table}
 \setlength{\tabcolsep}{10pt}
 \centering
 \caption{Impacts of adding Dollar General vaccine sites \label{tab:summaryrace}}
\begin{tabular}{    l c  c  c c }
\toprule
 & \multicolumn{2}{c}{Current partners} & \multicolumn{2}{c}{w Dollar General} \\
Population &  Share  &  Share   & Share  & Share \\
 type   &  $<$1 mile & $<$ 5 miles & $<$1 mile & $<$ 5 miles   \\ \midrule
All adults  & 48.3 & 86.3 & 56.3 & 94.3   \\
Black Population  & 53.6   &93.2  &66.1 & 98.0 	  \\
White Population  & 44.3  &83.8  & 52.3 & 93.3  \\
AAPI Population  &65.9 & 97.5 & 68.5 & 98.4 \\ 
Other Population & 58.5& 90.4& 65.0& 94.7\\
Hispanic Population & 44.9 &84.6  &53.4 & 93.7  \\
Non-Hispanic Population &61.2 & 93.7 & 68.0 & 97.0\\
\bottomrule
\end{tabular}
\begin{flushleft}
Summary statistics for Federal Retail Pharmacy Partner Program. Table entries denote the share of the total U.S. population type within the denoted proximity to an in-state pharmacy partner. 
\end{flushleft}
\end{table}

A few patterns are notable. First, while we find that a higher share of the Black population is within 1 mile of a pharmacy partner than the white population, the improvement in proximity of adding Dollar General outlets as vaccine sites is particularly large for the Black population. Adding Dollar General to the program would improve the fraction of Black individuals within a mile of a partner site from 53.6 percent to 66.1 percent. The improvement in proximity from adding Dollar General to the program is also disproportionately large for the Hispanic population relative to the non-Hispanic population. The share of Hispanic people less than a mile from a partner site is 44.9 percent without Dollar General and 53.4 percent with Dollar General. 

\begin{figure}[!ht]
\caption{Percentage of Low-Income Households with Federal Pharmacy Partner at <1mi}
\begin{center}
\includegraphics[trim={0 .15cm 0 .15cm},width=172mm]{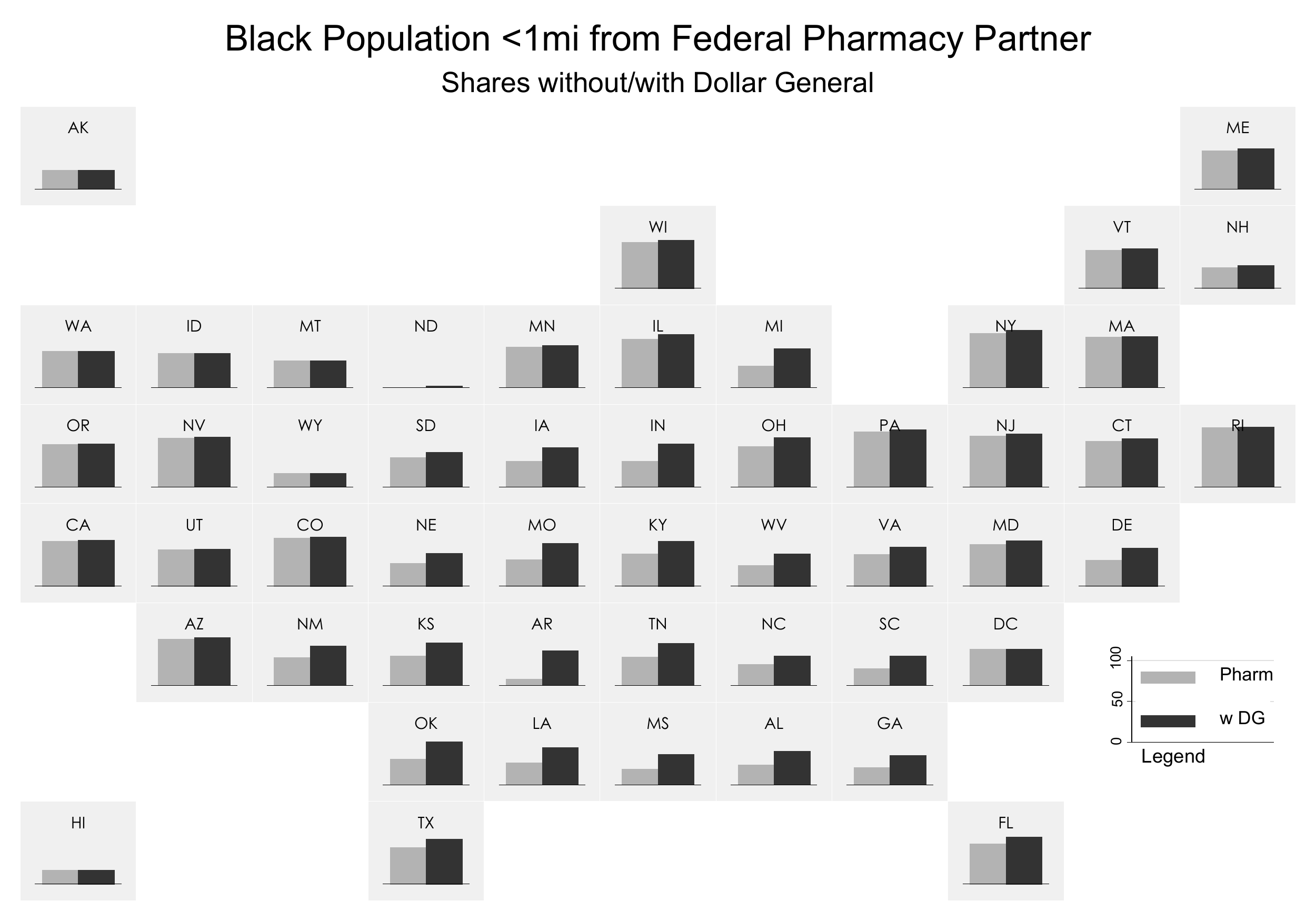}
\end{center}
\label{fig:blackpop} 
     \fnote{\small  Notes: State-by-state data on the share of households earning less than \$35K per year that are located less than a mile from a federal pharmacy partner. The grey bar represents the current pharmacy partners and the black bars add Dollar General as a partner. }
\end{figure}

We have seen that the overall data for the impacts of adding Dollar General to the program is particularly large for Black individuals and Hispanic people. For Black individuals, the overall data again masks substantial heterogeneity across states. As shown in Figure \ref{fig:blackpop}, the improvement to proximity for Black Americans is particularly pronounced throughout the Southeast and Midwest. Arkansas is one of the starkest examples. There, the federal pharmacy program outlets are within a mile of only 10.0 percent of blacks and 11.2 percent of whites. However, the addition of Dollar General improves one-mile proximity to 32.3 percent of whites and 53.6 percent of Black individuals. The benefit to black Americans of adding Dollar General is large in several Midwestern states.  For example, in Michigan, 28.2 percent of the white population is within one mile of a federal pharmacy partner and 33.6 percent of the black population. Adding Dollar General raises this to 37.5 percent of the white population and 60.2 percent of the Black population. The raw data underlying Figure \ref{fig:blackpop} is given in the Appendix. A similar data table for the Hispanic population is also given in the Appendix (\fref{tab:pharm_hisp_tbl}). 

Dollar General and the CDC have confirmed that a vaccine partnership between them is being discussed. However, given the high-SVI locations of Dollar Tree, it could be that Dollar Tree as a vaccine partner would provide even better proximity to vaccines for low-income households than Dollar General.\footnote{Of course, there may be other characteristics of Dollar Tree that make them less suitable or even totally unsuitable for the program, such as available space, etc.}  We examine this question by again measuring the proximity of low-income households to current pharmacy partners and to Dollar Tree. We found that 60.5 percent of low-income households in the continental U.S. are less than one mile from current pharmacy partners plus Dollar General while 61.56 percent of low-income households are less than one mile from current pharmacy partners plus Dollar Tree. Thus, Dollar Tree would provide slightly higher within one mile access to low-income households.  Interestingly, this finding is not true for wider distance bands.  More low-income households are within 2 or 5 miles from the current pharmacies plus Dollar General than are within 2 or 5 miles from the current pharmacies plus Dollar Tree.  

\begin{figure}[!ht]
\caption{Percentage of Low-Income Households with Federal Pharmacy Partner at <1mi}
\begin{center}
\includegraphics[trim={0 .15cm 0 .15cm},width=172mm]{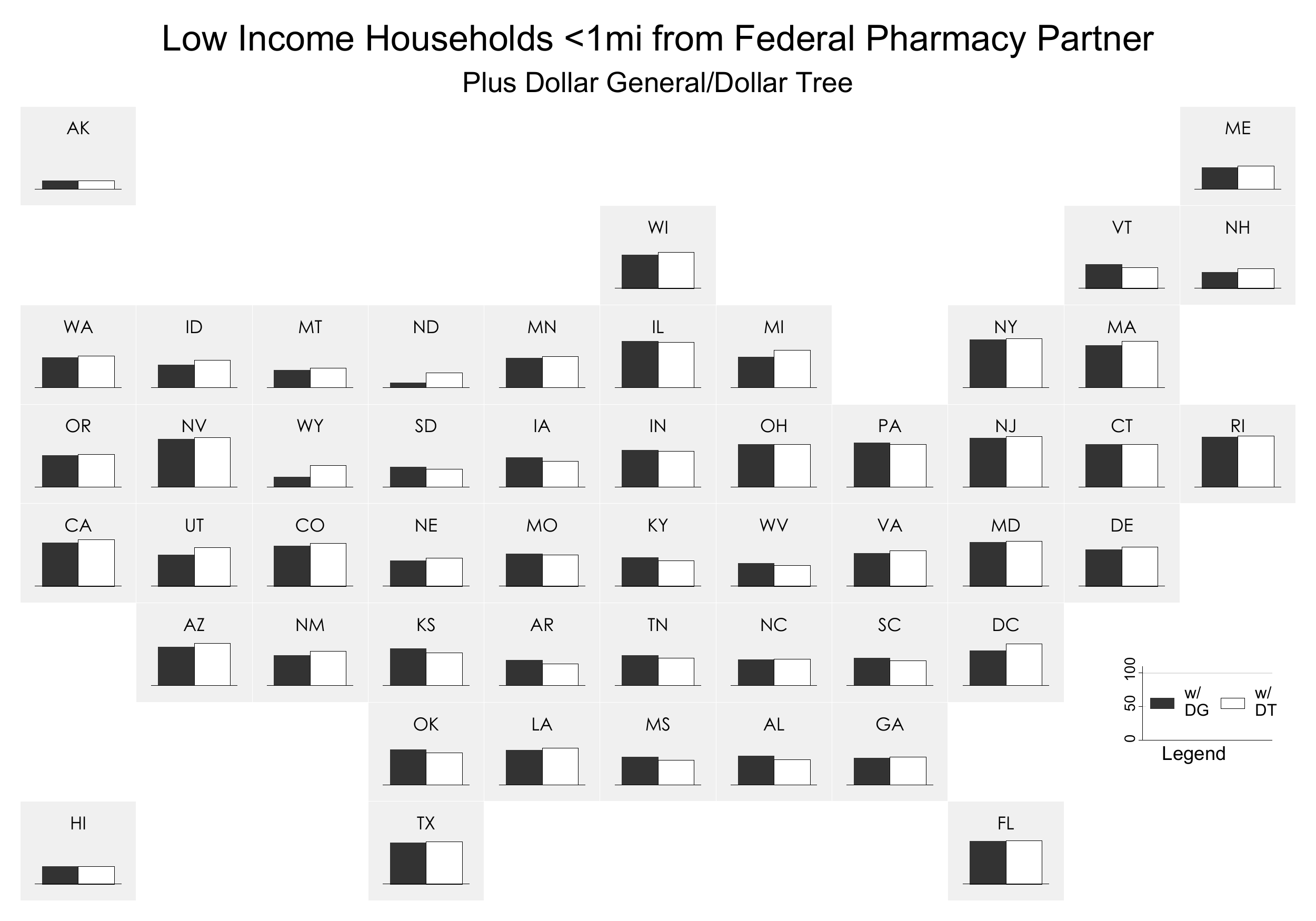}
\end{center}
\label{fig:distimprovementstateDTDG} 
     \fnote{\small  Notes: State-by-state data on the share of households earning less than \$35K per year that are located less than a mile from a federal pharmacy partner or dollar store. The black bars represents the current pharmacy partners plus Dollar General while the white bars represent the current pharmacy partners plus Dollar Tree. }
\end{figure}

The overall findings again mask some cross-state heterogeneity.  As Figure \ref{fig:distimprovementstateDTDG} shows, the share of low-income households within a mile of a pharmacy partner plus hypothetical dollar store partner is, in many states, similar whether the dollar store partner is Dollar General or Dollar Tree. Dollar Tree's locations are particularly attractive relative to Dollar General in the West.

\section{State Allocations and State Vaccination Sites}

Our findings that dollar stores vaccine sites would substantially expand vaccine proximity for low-income households, particularly in the South and Southeastern United States. Thus far, our analysis has examined only chains that are partners in the Federal Retail Pharmacy Partnership Program.  However, as discussed above, while some vaccine doses are allocated to this program, doses are also allocated to states to supply the vaccination sites that states support. If state vaccination sites have been established in low-income areas, it is possible that dollar stores and the sites chosen by states for vaccinations are redundant in increasing proximity to low-income residents. 

To do this, we scraped the COVID vaccine websites of 21 states to find the full listing of vaccination sites in the state. These listings typically included federal pharmacy partners and Federally Qualified Health Centers.  However, they typically included only sites that were actively providing vaccinations when we scraped the site. We used these listings to construct, for these 21 states, a superset of vaccine sites.  These included the federal retail partners that we had identified plus the sites listed in the state website. Obviously, sometimes a site was listed in both places. When a retail chain in the pharmacy program had not yet started using a particular outlet for vaccines, it would appear on our federal retail partner list but not on the state list. A state hospital or mass vaccination site would appear on the state list but not our federal partner list. Our superset of sites includes all of the sites on the state site plus all outlets of federal retail pharmacy partners.  Having constructed that set, we recalculated the distance of households to their closest within-state vaccination site, again the Census methodology described above. 

We find that state vaccination sites do not add substantially to our measures of proximity to low-income households in some states, but do in others. For example, we find that only 9.1 percent of low-income households were within a mile of the federal pharmacy partners in Arkansas. Adding in the state sites raises the fraction of low-income households within a mile of a vaccine site to 32 percent. In states for which the state sites do not meaningfully improve proximity, we hypothesize that either the state is using pop-up clinics not listed on the state site and/or the state relies heavily on mass vaccination sites that have large capacity but do not substantially improve proximity. While the state sites substantially improve proximity to low-income households, we find that in many cases, particularly in the South, the state sites are not redundant with the proximity benefits of adding dollar stores.  Not considering state sites, the combination of federal pharmacy partners and Dollar General outlets raised the fraction of Arkansas low-income households within a mile of a vaccine site from 9.1 percent without Dollar General to 40 percent with Dollar General.  The Federal pharmacy plus state sites are within a mile of 32 percent of low-income households. Adding Dollar General would still improve this share substantially, to 48 percent.  

Table \ref{tab:statescrape} shows the share of low-income households within a mile of a federal pharmacy partner, the share within a mile of a federal pharmacy partner and/or state site, and the share within a mile of a federal pharmacy partner, state site, and/or Dollar General. The figure shows that Dollar General provides substantial improvement in access, particularly in the South, even when taking state vaccine sites into account.  The table also shows the share of low-income households within five miles of a vaccine site using these same metrics.

\begin{table}[!ht]
 \setlength{\tabcolsep}{10pt}
 \centering
\caption{Characteristics of federal pharmacy and state vaccine sites  \label{tab:statescrape}}
\begin{threeparttable}
\small
    \begin{tabular}{l|ccc|ccc}
    \toprule
     & $<$1 mile  & $<$1 mile & $<$1 mile & $<$5 miles  & $<$5 miles & $<$5 miles   \\ 
            State  & pharm.  & pharm. + & pharm. + & pharm.  & pharm. + & pharm. + \\    
            &   &  state   & state + &   &  state   &state+\\   
            & &  & DG &    &    & DG \\    \midrule
Alaska&13.79&37.97&37.97&39.41&58.66&58.66\\
Alabama&26.79&36.36&51.98&69.6&76.98&93.58\\
Arkansas&9.12&32.04&48.43&50.24&73.89&87.63\\
Connecticut&64.68&73.21&75.54&97.33&98.01&98.72\\
Georgia&25.21&32.81&47.56&77.91&84.29&95.97\\
Illinois&66.15&71.27&77.89&89.33&90.7&96.38\\
Kansas&39.9&50.12&62.27&73.46&83.9&90.47\\
Maryland&64.53&72.11&76.02&92.97&94.41&97.34\\
Maine&32.45&33.56&35.75&62.3&65.12&73.89\\
Missouri&31.21&58.61&65.48&73.94&85.49&92.37\\
Mississippi&23.14&39.16&50.86&58.56&68.81&87.16\\
North Dakota&0&50.81&51.4&14.64&72.6&74.11\\
New Jersey&75.75&79.4&82.08&98.71&99.12&99.73\\
Nevada&75.25&78.24&80.22&91.74&92.31&93.16\\
Ohio&55.62&67.24&75.3&90.22&92.29&97.29\\
Oklahoma&30.52&43.29&62.87&72.36&78.41&92.85\\
Pennsylvania&64.49&70.31&74.69&90.59&92.86&97.76\\
South Carolina &27.38&32.87&47.58&76.56&79.02&94.13\\
Washington &48.27&56.07&56.18&83.96&88.75&88.85\\
West Virginia&22.45&23.1&37.63&68.4&70.11&87.03\\
\bottomrule
        \end{tabular}
       
\begin{tablenotes}
\item Notes: Share of low-income households less than one mile/5 miles from pharmacy partners, pharmacy partners plus state vaccine sites, and pharmacy partners plus state vaccine sites plus Dollar General. Data from RefUSA and state vaccine sites. 
\end{tablenotes}
\end{threeparttable}
\end{table}

\section{Discussion and Conclusions}

In the four months since its launch, the U.S. COVID-19 vaccination program has rapidly transitioned from having one authorized vaccine, extremely scarce doses available, and narrow vaccine prioritization guidelines to three authorized vaccines, tens of millions of doses available weekly, and open eligibility in many U.S. states. As this effort transitions to a new phase throughout the spring and summer, considerable work remains delivering vaccines to those yet to be vaccinated, for whom navigating vaccine appointment registration systems, identifying available vaccine sites, arranging for transportation, or other access barriers may have thus far been insurmountable obstacles. 

Evidence to date suggests that vaccination efforts
---nationally, at the state level, and in cities---have performed considerably less well with respect to meeting their equity goals as they have regarding those measuring the speed and pace of vaccine administration \citep{hughes2021county}. These reports also indicate that the need for enhanced efforts to ensure broad access to vaccines---and correspondingly high vaccine uptake---in low-income and high-SVI communities remains particularly acute. Even if vaccines were to be made available in some traditional clinical settings later in 2021, centrally-located, vaccination-specific sites in communities will most likely remain an important component of the COVID-19 response for months to come. This is particularly true given the possibility that emerging viral variants or waning immunity, two areas being monitored by health officials, may require subsequent cycles of mass vaccination that would require elements of the large-scale vaccination infrastructure and capacity currently in place and actively being expanded upon \citep{del2021covid}.

We show that adding dollar stores as vaccination sites, as has been recently discussed by the Biden administration, would offer considerable proximity benefits, particularly for low-income households and Black Americans in several regions of the continental U.S., if such a program were added to the existing retail pharmacy partnership. These benefits would be seen even if all available pharmacy partner locations were to serve as vaccination sites. Benefits from the addition of Dollar General stores, the specific partner in discussions with federal officials regarding a potential partnership, would be enhanced in many locations by the inclusion of Dollar Tree, the other major dollar store chain.

The use of dollar stores as vaccination sites would require a number of implementation and logistical challenges to be addressed, particularly because they currently lack the facilities and employees trained to administer vaccines present at federal pharmacy partner locations. But recent discussions about a potential dollar store partnership by federal health officials suggest that they believe those challenges to be solvable, offset by potential advantages including not only the well-positioned locations of these stores but also their familiarity to local residents, available indoor and outdoor space, parking lots (in many locations), and other attributes. Some of the \$7.5 billion for COVID-19 vaccination provided in the recently signed American Rescue Plan could be used for the vaccine-related equipment and personnel needed for dollar stores to serve as vaccination sites. 

If implemented successfully, the administration of vaccines at dollar stores would not only provide the proximity benefits considered invaluable to the equitable rollout of COVID-19 vaccines, but it could suggest additional opportunities beyond the current pandemic. Annual influenza vaccination similarly relies on rapidly delivering vaccines to tens of millions of Americans in a compressed period, often through the use of temporary large-scale vaccination clinics outside of traditional health care settings. Reported rates of influenza vaccine administration are lower among low-income older adults, the age group---as with COVID-19---at the greatest risk of severe disease-related outcomes \citep{artiga2020racial}.

Enhancing proximity to vaccination sites is by no means  sufficient to ensuring access, equity, and high rates of COVID-19 vaccination among low-income and minority communities or the overall population. But it is an important component of approaches to reduce barriers to vaccination, alongside efforts to build support and enthusiasm for the vaccines and address sources of vaccine hesitancy. The potential addition of dollar stores as vaccination sites---in tandem with the federal pharmacy program and state-supported locations---is a promising approach toward strengthening COVID-19 vaccination programs and increasing their ability to reach all eligible populations in the months ahead.

\clearpage
\section{References}
\setstretch{1}
\renewcommand{\bibsection}{}
\bibliographystyle{em}

{\raggedright
\bibliography{papers}
}

\clearpage
\section{Appendix}
\appendix
\setstretch{1.7}

\begin{table}[!ht]
 \setlength{\tabcolsep}{10pt}
 \centering
\caption{State-Coordinated Vaccine Locations \label{state_links}}
\begin{threeparttable}
\begin{tabular}{ll}
\toprule
State	&	Website	\\
\midrule
Alabama & {\scriptsize https://bamatracker.com/providers}\\
Alaska & \scriptsize https://anchoragecovidvaccine.org/providers/ \\
Arkansas &  {\scriptsize https://www.healthy.arkansas.gov/programs-services/topics/covid-19-map-of-1-a-pharmacy-locations}\\
Connecticut & \scriptsize https://www.211ct.org/search?page=1\&location=Connecticut\&taxonomy\_code=11172\&service\_area=connecticut\\
Georgia & \scriptsize http://www. dph.georgia.gov/locations/covid-vaccination-site\\
Illinois &\scriptsize https://coronavirus.illinois.gov/s/vaccination-location \\
Kansas &\scriptsize https://kdhe.maps.arcgis.com/apps/instant/nearby/index.html?appid=2cf619afb6c74320a26855840a8ca3e3\\
Maine &\scriptsize https://www.maine.gov/covid19/vaccines/vaccination-sites\\
Maryland &\scriptsize https://maryland.maps.arcgis.com/apps/instant/nearby/index.html?appid=0dbfb100676346ed9758be319ab3f40c\&find=\\
Mississippi &\scriptsize https://msdh.ms.gov/msdhsite/\_static/14,0,420,976.html\#providerMap\\
Missouri &\scriptsize https://covidvaccine.mo.gov/map/Approved-Vaccinators.pdf\\
North Dakota &\scriptsize https://app.powerbigov.us/view?r=eyJrIjoiNmY1ZWFiMzktYzMzNC00ZTQxLTkxZTAtNWRiMzkyYzYzMjk0IiwidCI6\\
&\scriptsize IjJkZWEwNDY0LWRhNTEtNGE4OC1iYWUyLWIzZGI5NGJjMGM1NCJ9\\
Nevada &\scriptsize https://www.immunizenevada.org/covid-19-vaccine-locator\\
New Jersey &\scriptsize https://newjersey.github.io/vaccine-locations/NJ-COVID-19-Vaccine-Locations.pdf\\
Ohio &\scriptsize https://coronavirus.ohio.gov/wps/portal/gov/covid-19/dashboards/covid-19-vaccine/covid-19-vaccine-provider-dashboard\\
Oklahoma &\scriptsize https://vaccinate.oklahoma.gov/en-US/vaccine-centers/\\
Pennsylvania &\scriptsize https://padoh.maps.arcgis.com/home/item.html?id=d169e1d2ae454bec928d046156dd7186\\
South Carolina &\scriptsize https://sc-dhec.maps.arcgis.com/apps/instant/nearby/index.html?appid=514e64ead13e4f508147dad8f483da38 \\
Washington &\scriptsize https://www.doh.wa.gov/YouandYourFamily/Immunization/VaccineLocations\#\\
West Virginia &\scriptsize https://dhhr.wv.gov/News/2021/Pages/COVID-19-Vaccination-Clinics-March-2-7,-2021.aspx\\
Wisconsin &\scriptsize https://dhsgis.wi.gov/server/rest/services/DHS\_COVID19/COVID19\_Vaccine\_Provider\_Sites/MapServer/0/query?\\
&\scriptsize where=1\%3D1\&text=\&objectIds=\&time=\&geometry=\&geometryType=esriGeometryEnvelope\&inSR=\&\\
&\scriptsize  spatialRel=esriSpatialRelIntersects\&relationParam=\&outFields=*\&returnGeometry=false\&returnTrueCurves=false\\
&\scriptsize  \&maxAllowableOffset=\&geometryPrecision=\&outSR=\&returnIdsOnly=false\&returnCountOnly=false\&orderByFields=\&\\
&\scriptsize  groupByFieldsForStatistics=\&outStatistics=\&returnZ=false\&returnM=false\&gdbVersion=\&returnDistinctValues=false\&\\
&\scriptsize  resultOffset=\&resultRecordCount=\&queryByDistance=\&returnExtentsOnly=false\&datumTransformation=\&\\
&\scriptsize  parameterValues=\&rangeValues=\&f=pjson\\
\bottomrule
\end{tabular}
\begin{tablenotes}
\item Notes: Websites used to gather state-coordinated vaccine locations. 
\end{tablenotes}
\end{threeparttable}
\end{table}

\clearpage
\begin{table}
 \setlength{\tabcolsep}{10pt}
 \centering
 \scriptsize
\caption{Summary Statistics for U.S. Dollar Stores, Selected Years \label{tab:dollarstores}}
\begin{threeparttable}
       \begin{tabular}{lllcc}
       \toprule
       & Dollar & Dollar & Dollar General & Dollar Tree \\ 
           & General & Tree & per 100K & per100K \\ \midrule
 Alabama&791&317&16.26&6.52\\
Alaska&0&0&0&0\\
Arizona&123&306&1.77&4.4\\
Arkansas&463&198&15.48&6.62\\
California&229&781&.58&1.99\\
Colorado&56&234&1.01&4.23\\
Connecticut&69&126&1.93&3.52\\
Delaware&48&65&5.06&6.85\\
District of Columbia&0&6&0&.88\\
Florida&927&1160&4.5&5.63\\
Georgia&944&719&9.17&6.98\\
Hawaii&0&0&0&0\\
Idaho&0&95&0&5.63\\
Illinois&587&547&4.58&4.27\\
Indiana&596&371&8.98&5.59\\
Iowa&276&110&8.81&3.51\\
Kansas&250&121&8.59&4.16\\
Kentucky&596&346&13.42&7.79\\
Louisiana&586&453&12.57&9.71\\
Maine&59&103&4.43&7.73\\
Maryland&138&237&2.3&3.95\\
Massachusetts&54&250&.79&3.66\\
Michigan&588&648&5.91&6.51\\
Minnesota&170&190&3.08&3.44\\
Mississippi&549&249&18.37&8.33\\
Missouri&564&302&9.26&4.96\\
Montana&0&40&0&3.84\\
Nebraska&130&70&6.83&3.68\\
Nevada&21&113&.72&3.87\\
New Hampshire&40&71&2.98&5.28\\
New Jersey&156&317&1.76&3.57\\
New Mexico&106&190&5.07&9.08\\
New York&499&662&2.54&3.37\\
North Carolina&896&743&8.82&7.32\\
North Dakota&44&36&5.85&4.79\\
Ohio&887&552&7.62&4.74\\
Oklahoma&472&241&12.05&6.15\\
Oregon&60&100&1.47&2.45\\
Pennsylvania&806&654&6.3&5.11\\
Rhode Island&21&73&1.99&6.91\\
South Carolina&575&396&11.6&7.99\\
South Dakota&60&42&6.94&4.86\\
Tennessee&832&403&12.51&6.06\\
Texas&1591&1732&5.71&6.21\\
Utah&11&126&.36&4.14\\
Vermont&38&24&6.08&3.84\\
Virginia&441&459&5.24&5.46\\
Washington&3&140&.04&1.92\\
West Virginia&240&191&13.12&10.44\\
Wisconsin&210&275&3.63&4.76\\
Wyoming&3&45&.52&7.73\\
\bottomrule
         \end{tabular} 
         
\begin{tablenotes}
\item Notes: Number of dollar stores and stores per 100 thousand population, by state. Data from RefUSA. 
\end{tablenotes}
\end{threeparttable}

\end{table}

\clearpage
\begin{table}
 \setlength{\tabcolsep}{10pt}
 \centering
 \scriptsize
\caption{Vaccine Access---Low-Income Households  \label{tab:pharm_inc_tbl}}
\begin{threeparttable}
    \begin{tabular}{lccc}
    \toprule
     & $<$1 mile  & $<$1 mile & $<$1 mile  \\ 
            State  & pharmacy  & pharmacy plus & pharmacy plus \\    
            &   &  DG   & DT  \\    \midrule
AK&13.79&13.79&13.79\\
AL&26.79&46.9&41.11\\
AR&9.12&40.72&35.31\\
AZ&57.43&62.87&68.73\\
CA&68.07&70.54&75.08\\
CO&61.29&65.38&69.58\\
CT&64.68&69.05&69.3\\
DC&56.56&56.56&67.3\\
DE&47.17&59.53&63.72\\
FL&61.17&69.83&70.23\\
GA&25.21&43.6&44.73\\
HI&28.99&28.99&28.99\\
IA&28.51&47.28&41.84\\
ID&36.7&36.7&44.84\\
IL&66.15&75.04&73.2\\
IN&34.25&59.97&57.58\\
KS&39.9&60.15&53.04\\
KY&30.54&46.91&41.38\\
LA&33.85&56.07&59.33\\
MA&67.16&68.61&75.03\\
MD&64.53&70.91&73.05\\
ME&32.45&34.65&37.41\\
MI&29.44&49.28&60.31\\
MN&42.34&47.67&50.43\\
MO&31.21&52.68&50.61\\
MS&23.14&44.85&40.4\\
MT&28.58&28.58&31.65\\
NC&28.7&42.44&42.97\\
ND&0&7.6&23.82\\
NE&20.42&41.46&45.31\\
NH&19.72&25.96&31.73\\
NJ&75.75&79.11&81.37\\
NM&32.91&48.43&55.49\\
NV&75.25&77.24&80.29\\
NY&71.69&77.85&79.62\\
OH&55.62&69.18&68.72\\
OK&30.52&57.38&52.22\\
OR&48.51&51.17&52.67\\
PA&64.49&71.52&68.89\\
PR&35.41&35.41&35.41\\
RI&78.53&80.53&82.48\\
SC&27.38&44.97&40.21\\
SD&18.73&32.22&28.89\\
TN&31.18&49.18&44.69\\
TX&51.61&66.81&68.41\\
US&48.39&59.58&60.65\\
UT&47.28&51&62.43\\
VA&42.92&53.26&57.26\\
VT&32.4&39.16&33.42\\
WA&48.27&48.38&51.62\\
WI&48.81&54.4&58.52\\
WV&22.45&37.47&33.71\\
WY&14.21&16.23&35.24\\
\bottomrule
         \end{tabular} 
        
\begin{tablenotes}
\item Notes: Share of low-income households less than one mile from pharmacy partners, pharmacy partners plus Dollar General, and pharmacy partners plus Dollar Tree/Family Dollar. Data from RefUSA.
\end{tablenotes}
\end{threeparttable}

\end{table}

\clearpage
\begin{table}
 \setlength{\tabcolsep}{10pt}
 \centering
 \scriptsize
\caption{Vaccine Access---Black Population \label{tab:pharm_black_tbl}}
\begin{threeparttable}
    \begin{tabular}{lccc}
    \toprule
     & $<$1 mile  & $<$1 mile & $<$1 mile  \\ 
            State  & pharmacy  & pharmacy plus & pharmacy plus \\    
            &   &  DG   & DT  \\    \midrule
AK&29.27&29.27&29.27\\
AL&30.74&52.02&51.04\\
AR&9.96&53.57&52.26\\
AZ&71.43&74.65&79.63\\
CA&69.93&71.25&77.09\\
CO&74.88&76.23&81.43\\
CT&70.39&74.85&74.83\\
DC&56.72&56.72&71.66\\
DE&40.19&59&60.98\\
FL&62.61&72.75&75.54\\
GA&27.2&45.23&48.75\\
HI&21.71&21.71&21.71\\
IA&39.57&61.12&65.09\\
ID&53.05&53.05&60.09\\
IL&75.37&82.61&84.54\\
IN&40.11&66.72&73.54\\
KS&45.99&65.78&68.56\\
KY&50&69.82&68.13\\
LA&34.41&57.68&63.69\\
MA&77.89&79.32&85.03\\
MD&64.45&70.53&74.09\\
ME&59.56&62.48&67.7\\
MI&33.61&60.18&81.93\\
MN&62.66&64.99&73.2\\
MO&41.09&66.62&78.57\\
MS&24.17&47.39&42.7\\
MT&41.69&41.69&41.99\\
NC&32.77&45.63&51.68\\
ND&0&2.66&40.05\\
NE&35.65&50.92&75.55\\
NH&32.54&35.63&45.51\\
NJ&79.01&82.46&85.2\\
NM&43.6&61.59&71.06\\
NV&75.98&77.36&81.46\\
NY&84.04&89.31&93.03\\
OH&62.93&76.51&79.73\\
OK&39.47&66.47&68.38\\
OR&65.46&66.29&70.04\\
PA&85.02&88.49&88.91\\
RI&92.02&92.61&93.13\\
SC&26.29&46.02&39.14\\
SD&45.43&53.63&63.85\\
TN&44.02&65.56&65.62\\
TX&56.41&69.41&72.98\\
UT&56.34&57.47&73.15\\
VA&49.39&60.92&66.45\\
VT&59.33&61.84&59.33\\
WA&55.98&55.99&60.51\\
WI&71.05&74.76&83.46\\
WV&32.06&49.74&48.66\\
WY&20.78&21.23&34.93\\
\bottomrule
         \end{tabular}
         
\begin{tablenotes}
\item Notes: Share of Black population less than one mile from pharmacy partners, pharmacy partners plus Dollar General, and pharmacy partners plus Dollar Tree/Family Dollar. Data from RefUSA.
\end{tablenotes}
\end{threeparttable}

\end{table}

\clearpage
\begin{table}
 \setlength{\tabcolsep}{10pt}
 \centering
 \scriptsize
\caption{Vaccine Access---Hispanic Population \label{tab:pharm_hisp_tbl}}
\begin{threeparttable}
    \begin{tabular}{lccc}
    \toprule
     & $<$1 mile  & $<$1 mile & $<$1 mile  \\ 
            State  & pharmacy  & pharmacy plus & pharmacy plus \\    
            &   &  DG   & DT  \\    \midrule
AK&15.72&15.72&15.72\\
AL&22.72&40.03&32.39\\
AR&10.65&35.34&28.59\\
AZ&59.54&62.15&64.04\\
CA&67.22&68.43&71.09\\
CO&61.99&63.59&66.34\\
CT&44.88&49&48.99\\
DC&63.99&63.99&73.04\\
DE&35.53&44.95&48.58\\
FL&52.56&60.93&59.62\\
GA&24.05&38.49&36.72\\
HI&24.27&24.27&24.27\\
IA&27.36&42.72&37.47\\
ID&35.99&35.99&43.4\\
IL&59.5&66.92&64.04\\
IN&29.62&48.57&42.41\\
KS&40.97&54.72&48.1\\
KY&28.78&44.76&38.88\\
LA&31.53&49.66&50.29\\
MA&52.73&54.14&58.55\\
MD&53.94&58.54&59.61\\
ME&31.15&34.47&36.94\\
MI&29.05&42.58&48.34\\
MN&39.77&43.32&45.76\\
MO&31.26&47.97&43.53\\
MS&19.23&36.8&31.39\\
MT&30.5&30.5&33.27\\
NC&26.64&37.6&37.35\\
ND&0&5.18&29.43\\
NE&26.76&44.69&42.88\\
NH&11.57&17.83&21.88\\
NJ&63.84&66.25&68.18\\
NM&34.37&47.2&51.12\\
NV&63.5&65.46&66.94\\
NY&64.36&68.92&69.88\\
OH&46.58&57.87&53.75\\
OK&30.7&53.01&44.89\\
OR&47.86&50.26&51.95\\
PA&51.56&58.32&54.65\\
RI&61.12&62.96&65.06\\
SC&26.17&42.23&35.01\\
SD&22.56&35.46&30.69\\
TN&28.14&43.9&36.87\\
TX&52.64&62.21&60.78\\
UT&40.51&41.96&52.06\\
VA&46.78&53.21&55.1\\
VT&32.35&38.38&33.21\\
WA&44.52&44.64&47.48\\
WI&35.75&41.6&44.23\\
WV&17.96&31.97&27.25\\
WY&14.16&15.52&31.17\\

\bottomrule
         \end{tabular}
         
\begin{tablenotes}
\item Notes: Share of Hispanic population less than one mile from pharmacy partners, pharmacy partners plus Dollar General, and pharmacy partners plus Dollar Tree/Family Dollar. Data from RefUSA.
\end{tablenotes}
\end{threeparttable}

\end{table}

\clearpage
\begin{table}
 \setlength{\tabcolsep}{10pt}
 \centering
 \footnotesize
\caption{Vaccine Access---Alternative Partners and Distance \label{tab:pharm_robust}}
\begin{threeparttable}
    \begin{tabular}{lcccccc}
    \toprule
     & $<$1 mile  & $<$1 mile & $<$1 mile & $<$5 miles  & $<$5 miles & $<$5 miles   \\ 
            State  & pharmacy  & pharmacy plus & pharmacy plus & pharmacy  & pharmacy plus & pharmacy plus \\    
            &   &  state   & DG &   &  state   & DG   \\    \midrule
AK&13.79&37.97&37.97&39.41&58.66&58.66\\
AL&26.79&36.36&51.98&69.6&76.98&93.58\\
AR&9.12&32.04&48.43&50.24&73.89&87.63\\
CT&64.68&73.21&75.54&97.33&98.01&98.72\\
GA&25.21&32.81&47.56&77.91&84.29&95.97\\
IL&66.15&71.27&77.89&89.33&90.7&96.38\\
KS&39.9&50.12&62.27&73.46&83.9&90.47\\
MD&64.53&72.11&76.02&92.97&94.41&97.34\\
ME&32.45&33.56&35.75&62.3&65.12&73.89\\
MO&31.21&58.61&65.48&73.94&85.49&92.37\\
MS&23.14&39.16&50.86&58.56&68.81&87.16\\
ND&0&50.81&51.4&14.64&72.6&74.11\\
NJ&75.75&79.4&82.08&98.71&99.12&99.73\\
NV&75.25&78.24&80.22&91.74&92.31&93.16\\
OH&55.62&67.24&75.3&90.22&92.29&97.29\\
OK&30.52&43.29&62.87&72.36&78.41&92.85\\
PA&64.49&70.31&74.69&90.59&92.86&97.76\\
SC&27.38&32.87&47.58&76.56&79.02&94.13\\
WA&48.27&56.07&56.18&83.96&88.75&88.85\\
WV&22.45&23.1&37.63&68.4&70.11&87.03\\
\bottomrule
        \end{tabular}
        
\begin{tablenotes}
\item Notes: Share of low-income households less than one mile/5 miles from pharmacy partners, pharmacy partners plus state vaccine sites, and pharmacy partners plus state vaccine sites plus Dollar General.Data from RefUSA and state vaccine sites.
\end{tablenotes}
\end{threeparttable}

\end{table}

\end{document}